\documentclass[12pt,epsf]{article}
\usepackage{epsf,graphicx}
\def\be{\begin{equation}}
\def\ee{\end{equation}}
\def\bea{\begin{eqnarray}}
\def\eea{\end{eqnarray}}

\begin{document}

\begin{center}
{{\Large \bf{
Implications of the first  neutral
current data from SNO for Solar Neutrino Oscillation}}}
\vskip 2cm
Abhijit Bandyopadhyay$^a$\footnote{e-mail: abhi@theory.saha.ernet.in},
 Sandhya Choubey$^b$\footnote{email: sandhya@hep.phys.soton.ac.uk},
 Srubabati Goswami$^c$\footnote{e-mail: sruba@mri.ernet.in},
D.P. Roy$^{de}$\footnote{e-mail:dproy@theory.tifr.res.in}
\vskip 1cm
$^a$Saha Institute of Nuclear Physics, Bidhannagar, Kolkata
700 064, INDIA\\
$^b$
Department of Physics and Astronomy, University of Southampton, 
Highfield, Southampton S017 1BJ, UK\\
$c$
Harish-Chandra Research Institute, Chhatnag Road, Jhusi, 
Allahabad - 211-019, INDIA
$^d$Tata Institute of Fundamental Research, Homi Bhabha
Road, Mumbai 400 005, INDIA \\ 
$^e$Physics Department, University of California,
Riverside, Ca 92521,USA
\end{center}
 
\vskip 1cm
\begin{abstract}

We perform model independent and model dependent 
analyses of solar neutrino data including the neutral current event rate 
from SNO. The inclusion of the first SNO NC data 
in the model independent
analysis determines the  
allowed ranges of $^{8}{B}$ flux normalisation 
and the $\nu_e$ survival probability more precisely than what was possible from 
the SK and SNO CC combination.   
We perform global $\nu_e-\nu_{active}$ oscillation analyses 
of solar neutrino data 
using the NC rate instead of the SSM prediction for the $^{8}{B}$ flux,
in view of the large uncertainty in the latter. The LMA gives the best
solution, while the LOW solution is allowed only at the $3\sigma$ level. 
\end{abstract}

\newpage

The neutral current results from the Sudbury Neutrino Observatory 
measures for the first time the total flux of $^{8}{B}$ neutrinos coming 
from the Sun \cite{snonc}.  
In a recent paper \cite{bcgd} we had examined the role of the 
anticipated  NC data from 
SNO in enhancing our understanding of the solar neutrino problem. 
The SNO NC rate can be expressed 
in terms of SNO CC and SK elastic scattering rates 
as \cite{barger} 
\be
R^{NC}_{SNO} = R^{CC}_{SNO} + (R^{el}_{SK} - R^{CC}_{SNO})/r,
\label{one}
\ee
where $r = \sigma^{NC}_{\nu_\mu,\tau}/\sigma^{CC+NC}_{\nu_e} \simeq 0.157$
for a threshold energy of 5 MeV (including the detector resolutions and the 
radiative corrections to $\nu-e$ scattering cross-sections). 
All the rates 
are defined with respect to the BBP2000 
Standard Solar Model (SSM) \cite{bbp2000}. 
We showed in \cite{bcgd} that 
because SNO has a greater sensitivity  to the NC scattering rate 
as compared to SK, the SNO NC measurement will be 
more precise and hence incorporation of this can be more predictive than 
the SNO CC and SK combination. 
We took three representative NC rates 
-- $R_{NC}^{SNO}$ = 0.8,1.0 and 1.2 ($\pm 0.08$)  
and showed that 
\begin{enumerate}
\item
For a general transition of $\nu_e$ into a mixture of active and sterile
neutrinos the size of the sterile component can be better 
constrained than  before. 

\item 
For transition to a purely active neutrino the $^8{B}$ neutrino flux 
normalisation and the survival probability $P_{ee}$ 
are determined more precisely. 
 
\item
We had 
also performed global two flavour 
oscillation analysis of the solar neutrino 
data for the $\nu_e -\nu_{active}$ case, where
instead of $R_{SK}$ and $R^{CC}_{SNO}$ we  used the quantities 
$R^{el}_{SK}/R^{NC}_{SNO}$ and $R^{CC}_{SNO}/R^{NC}_{SNO}$. 
These ratios are independent of the  
$^8B$ flux normalisation and hence of the SSM uncertainty.  
We showed that use of  these ratios can result in drastic 
reduction of the allowed parameter regions specially in the LOW-QVO 
area  
depending on the value of the NC rate. 

\end{enumerate}

We now have the actual experimental result
\be
R^{NC}_{SNO} = 1.01 \pm 0.12
\ee
while
eq. (\ref{one}) gives $1.05 \pm 0.15$.
Thus in 306 live days (577 days)
the SNO NC measurement has achieved a precision, which is already
better than that obtained from the SK and SNO CC combination. 
This paper  follows closely the analysis that we have done in \cite{bcgd}
but incorporating the actual data. 
In addition we also perform an alternative global analysis  for 
$\nu_e -\nu_{active}$ oscillation
by letting the $^{8}{B}$ normalisation factor $f_B$ vary freely, 
where the inclusion of 
$R^{NC}_{SNO} ( = f_B)$ in the fit serves to control this parameter.  
As we shall see below the two methods of global analysis give very
similar results.

In section 1  we discuss the constraints on the 
electron neutrino survival probability, 
the $^{8}{B}$ normalisation factor $f_B$ and the fraction 
of sterile component without assuming any particular model for the 
probabilities. 
In section 2 we perform the global  analyses assuming 
two flavour $\nu_e - \nu_{active}$ oscillation.  

\section{Model Independent Analysis} 
For the general case of $\nu_e$ transition into a combination of 
$\nu_{active}$ ($\nu_a$) and $\nu_{sterile}$ ($\nu_s$) states 
one can write the 
SK, SNO CC and SNO NC rates as  
\bea
R^{el}_{SK} &=& f_B P_{ee} + f_B r P_{ea},
\label{two} \\[2mm]
R^{CC}_{SNO} &=& f_B P_{ee},
\label{three} \\[2mm]
R^{NC}_{SNO} &=& f_B (P_{ee} + P_{ea}),
\label{four}
\eea
where $P_{ee}$ and $P_{ea}$ denote the probabilities
folded with the detector response  
function \cite{villante}
and averaged over energy. 
To extract a model independent bound on $P_{ee}$ 
one has to ensure an equality of the response functions
which amounts to slight adjustment of the SK threshold energy and the 
rate 
\cite{villante,lisisno}. 
Our approach is slightly different. 
We treat  $P_{ee}$ to be effectively energy independent.
The SK spectrum data
indicates a flat probability down to 5 MeV \cite{sk}. This is corroborated 
by SNO \cite{sno,snodn} which now has a 
threshold of 5 MeV for kinetic energy of the observed electron.  
Hence we consider this assumption as justified 
and expect the results to be insensitive to the differences in
the response functions. 
It should be noted here that in cotrast to the SNO CC events their NC events
correspond to a neutrino energy threshold of 2.2 MeV. However it is
clear from eq. (5) that for a $\nu_e$ to $\nu_a$ transition there is no reason
to expect any energy dependence in $R^{NC}_{SNO}$. 
On the other hand for the general
case of $\nu_e$ transition into a combination of $\nu_a$ and 
$\nu_s$ our approach
effectively assumes $P_{es}$ to be energy independent down to 2.2 MeV.
A comparison of the current values   
$R^{CC}_{SNO}$ with $R^{NC}_{SNO}$ is shown in fig. 1. It constitutes 
a 5.3$\sigma$ signal for transition to a state containing an 
active neutrino component or alternatively a 5.3$\sigma$ signal against a 
pure sterile solution.  

Next we consider the general case where $\nu_e$ goes to
a mixed state = $\nu_{a} \sin \alpha + \nu_s \cos \alpha$.  
Then one can write $P_{ea} = \sin^2\alpha (1 -P_{ee}$).   
Substituting this in the equations (\ref{two}) and (\ref{four}) 
and eliminating $P_{ee}$ using equation (\ref{three}) one gets the following 
set of equations for $f_B$ and $\sin^2\alpha$ \cite{bcgd} 
\bea
\sin^2\alpha (f_B - R^{CC}_{SNO}) &=& (R^{el}_{SK} - R^{CC}_{SNC})/r,
\label{five} 
\\[2mm]
\sin^2\alpha (f_B - R^{CC}_{SNO}) &=& R^{NC}_{SNO} - R^{CC}_{SNO}.
\label{six}
\eea
We treat $\sin^2\alpha$ as a model parameter. And for different input 
values of $\sin^2\alpha$ we determine the central value and 
the $1\sigma$ and $2\sigma$ ranges of  $f_B$ by taking a weighted 
average of the equations (\ref{five}) and (\ref{six}). 
The corresponding curves are presented in fig. 2. Combining the
2 $\sigma$ lower limit of $f_B$ from this fit with the $2\sigma$ upper
limit from the SSM (vertical lines) gives a lower limit of 
$\sin^2\alpha$ $>$ 0.45 i.e. the probability of the active component
is $>$ 45\%. Note that there is no upper limit on this quantity - i.e.
the data is perfectly compatible with $\nu_e$ transition into purely
active neutrinos.

Assuming transition into purely active neutrinos ($P_{ea} = 1 - P_{ee}$ )
we show in fig. 3 the $1\sigma$ and $2\sigma$ contours in the $f_B -P_{ee}$
plane from the combinations SK+SNOCC 
and SK+SNOCC+SNONC.
The inclusion of the NC rate 
narrows down the ranges of $f_B$ and $P_{ee}$. The error in $f_B$ 
after the inclusion of NC data is about half the size of
the corresponding error 
from SSM as is seen from fig. 3.  

\section{Model dependent analysis} 

In this section we present the results of our $\chi^2$ analysis of solar 
neutrino rates and SK spectrum data in the framework of two flavour 
oscillation of $\nu_e$ 
to an active flavour. 
We use the standard techniques described in our earlier 
papers \cite{sg,bcgk} excepting for the fact that 
instead of 
the quantities  $R^{el}_{SK}$ and $R^{CC}_{SNO}$ we now fit the ratios 
$R^{el}_{SK}/R^{NC}_{SNO}$ and
$R^{CC}_{SNO}/R^{NC}_{SNO}$. 
The $^{8}{B}$ flux normalisation 
gets cancelled from these ratios and the analysis becomes 
independent of the  
large (16-20\%) SSM uncertainty  associated with this.       
We include in our global analyses the 1496 day SK zenith angle 
spectra \cite{smy}.
Since we use both SK rate and SK spectrum data 
we keep a free normalisation factor for the SK spectrum. 
This amounts to taking the information on total rates from the SK rates 
data and the information of the spectral shape from the SK spectrum data. 
The SNO CC and NC rates have a large anticorrelation. We have 
taken into account this correlation between the measured SNO rates 
in our global analyses.
Further details of this fitting method
can be found in \cite{bcgd}. 
In Table 2 we present the best-fit parameters, $\chi^2_{min}$ 
and goodness of fit (GOF). 
The best-fit comes in the HIGH(LMA) region as before \cite{bcgk,lma}. 
However as is seen from fig. 4a the incorporation of the 
NC data narrows down the allowed regions, and in particular  
the  LOW region becomes much smaller. 

We have also performed an alternative $\chi^2$ fit to the rates of Table 1
\cite{snonc,sk,ga,cl}
along with the 1496 day SK spectra \cite{smy},
keeping $f_B$ as a free parameter. 
Even though we allow $f_B$ to vary freely the NC data serves to control $f_B$ 
within a range determined by its error. As we see from Table 2 and fig. 4b 
the results of this fit are very similar to the previous case. The best
fit comes from the HIGH(LMA) region, while no allowed region is obtained 
for the LOW solution 
at the 99\% CL level. Maximal mixing is seen to be 
disallowed at the $3\sigma$ level. 
To illustrate the impact of the NC rate
on the oscillation solutions we have repeated the free $f_B$ fit
without this rate. The results are shown in fig 4c.  
Evidently the NC data plays a pivotal role in
constraining the oscillation solutions, particularly in the
LOW/QVO region, which is allowed only at the $3\sigma$ level. 
It puts an upper bound on the $\Delta m^2$ in the LMA 
region and rules out maximal mixing. 

\section{Summary and Conclusions}

The first SNO NC data constitutes a 5.3$\sigma$ signal 
for transition into a state containing an active neutrino component. 
The inclusion of this data puts much  tighter constraints on $f_B$ and $P_{ee}$ 
from a model independent analysis involving active neutrinos 
as compared to the 
SNO CC/SK combination. 
In this paper we have discussed two useful strategies, of incorporating
the NC data in the global $\chi^2$ analysis of rates and spectrum data, 
by which one can avoid the large 
$^{8}{B}$ flux uncertainty from the SSM. \\
$\bullet$ We fit the ratios of the SK elastic and SNO CC rates w.r.t the NC rate, from which the $f_B$ cancels out.  \\
$\bullet$ We fit the rates by keeping $f_B$ as a free parameter, where the  inclusion of the SNO NC rate (= $f_B$) serves to control this parameter. 

Both the analyses give very similar results. They clearly favour the 
HIGH(LMA) solution, while a limited region of the LOW solution is also
acceptable at the $3\sigma$ level. The maximal mixing solution is disfavoured 
at the $3\sigma$ level.
As more data accumulate
one expects a substantial reduction in the error bar of the SNO NC rate,
resulting in further tightening of the allowed regions of neutrino mass
and mixing.
\\ \\
{\it Note Added}: The paper \cite{bargernc} appeared on the 
net after completion
of our work. In the region of overlap our results agree
with theirs as well as 
with the updated version of \cite{strumia}. 
It may be added here that the SNO  CC and NC rates given in Table 1 are 
obtained
assuming undistorted energy spectra above 5 MeV, which 
for transitions to active neutrinos has good
empirical
justification as mentioned above.
We thank Prof. Mark 
Chen of SNO collaboration for
communication on this point.

\newpage
\begin{table}
\begin{center}
\begin{tabular}{|c|c|c|}
\hline
&& \\
experiment & $R$ & composition  \\
&& \\
\hline
&& \\
$Ga$ & 0.553 $\pm$ 0.034 & $pp(55\%),Be(25\%),B(10\%)$  \\
&& \\
$Cl$ & 0.337 $\pm$ 0.030 & $B(75\%),Be(15\%)$  \\
&& \\
$SK$ & 0.465 $\pm$ 0.014 (0.363 $\pm$ 0.014) & $B(100\%)$ \\
&& \\
$SNO(CC)$ & 0.349 $\pm$ 0.021 & $B(100\%)$ \\
&& \\
$SNO(NC)$ & 1.008 $\pm$ 0.123 & $B(100\%)$ \\
&& \\
\hline
\end{tabular}
\caption{
The observed solar neutrino rates relative to the $SSM$
predictions (BP2000) are shown along with their compositions 
for different experiments.  For the $SK$ experiment the
$\nu_e$ contribution to the rate $R$ is shown in parantheses assuming
$\nu_e \rightarrow \nu_a$ transition. In the combined Ga rate we have 
included the latest data from SAGE and GNO.}
\end{center}
\end{table}

\begin{table}
\begin{center}
\begin{tabular}{cccccc}
\hline
Data&Nature of & $\Delta m^2$ &
$\tan^2\theta$&$\chi^2_{min}$& Goodness\\
Used&Solution & in eV$^2$&  & & of fit\\
\hline
Ga + &LMA & $9.66 \times 10^{-5}$&$0.41$ & 35.95 & 80.08\%\\
SK/NC + &LOW & $1.04\times 10^{-7}$ & 0.61 & 46.73 & 36.09\%\\
CC/NC + &VO& $4.48\times 10^{-10}$& 0.99 & 54.25  & 13.84\%\\
SKspec&SMA& $6.66\times 10^{-6}$ & $1.35\times 10^{-3}$ & 67.06 & 1.41\%\\
\hline
Cl + Ga +&LMA & $6.07 \times 10^{-5}$&$0.41$ & 40.57 & 65.99\%\\
SK + CC +&LOW & $1.02\times 10^{-7}$ & 0.60 & 50.62 & 26.14\%\\
NC + SKspec &VO& $4.43\times 10^{-10}$& 1.1 & 56.11  & 12.39\%\\
+ $f_B$ free&SMA& $5.05\times 10^{-6}$ & $1.68\times 10^{-3}$ & 70.97 & 0.81\%\\\hline
\end{tabular}
\caption
{The $\chi^2_{min}$, the goodness of fit
and the best-fit values of the oscillation
parameters obtained for the analysis of the global solar neutrino
data.}
\end{center}
\end{table}

\newpage
\begin{figure}
\begin{center}
\hspace{10cm}
\epsfxsize=5in
\epsfbox{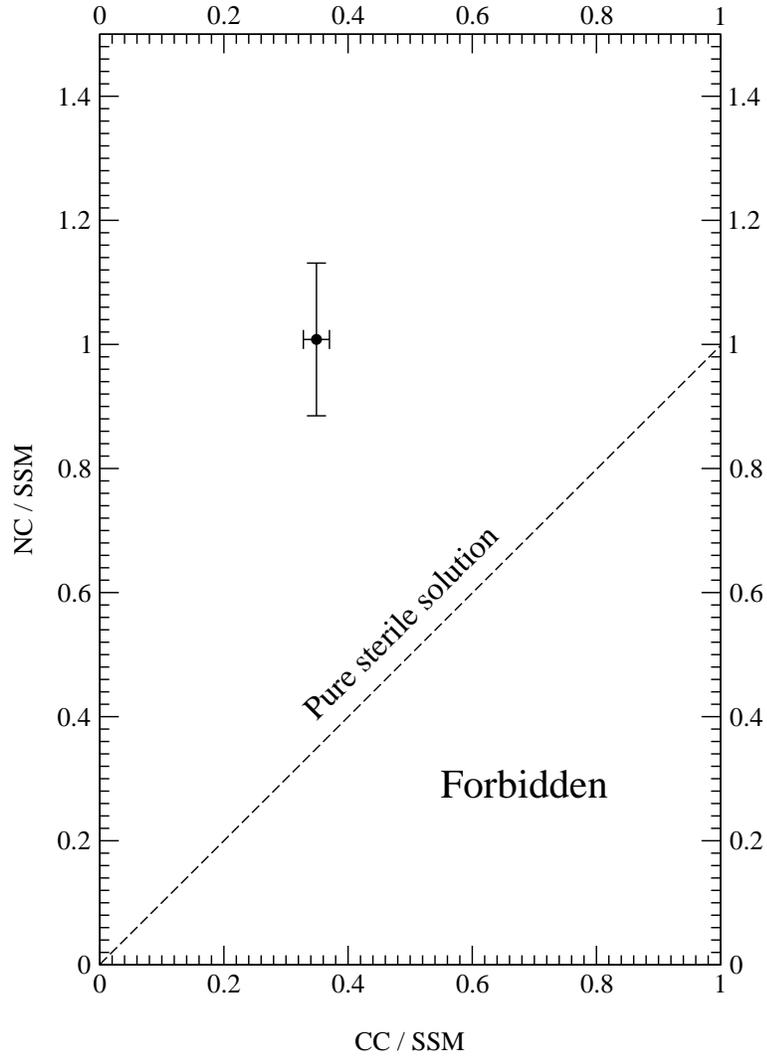}
\label{fig:ncfig1}
\caption{The $SNO$ $CC$ and $NC$ rates shown relative to their $SSM$
predictions. The
dashed line is the prediction of the pure $\nu_e$ to $\nu_s$ transition.
The pure sterile solution is seen to be disfavored at $5.3 \sigma$.}
\end{center}
\end{figure}

\newpage

\begin{figure}
\begin{center}
\hspace{10cm}
\epsfxsize=5in
\epsfbox{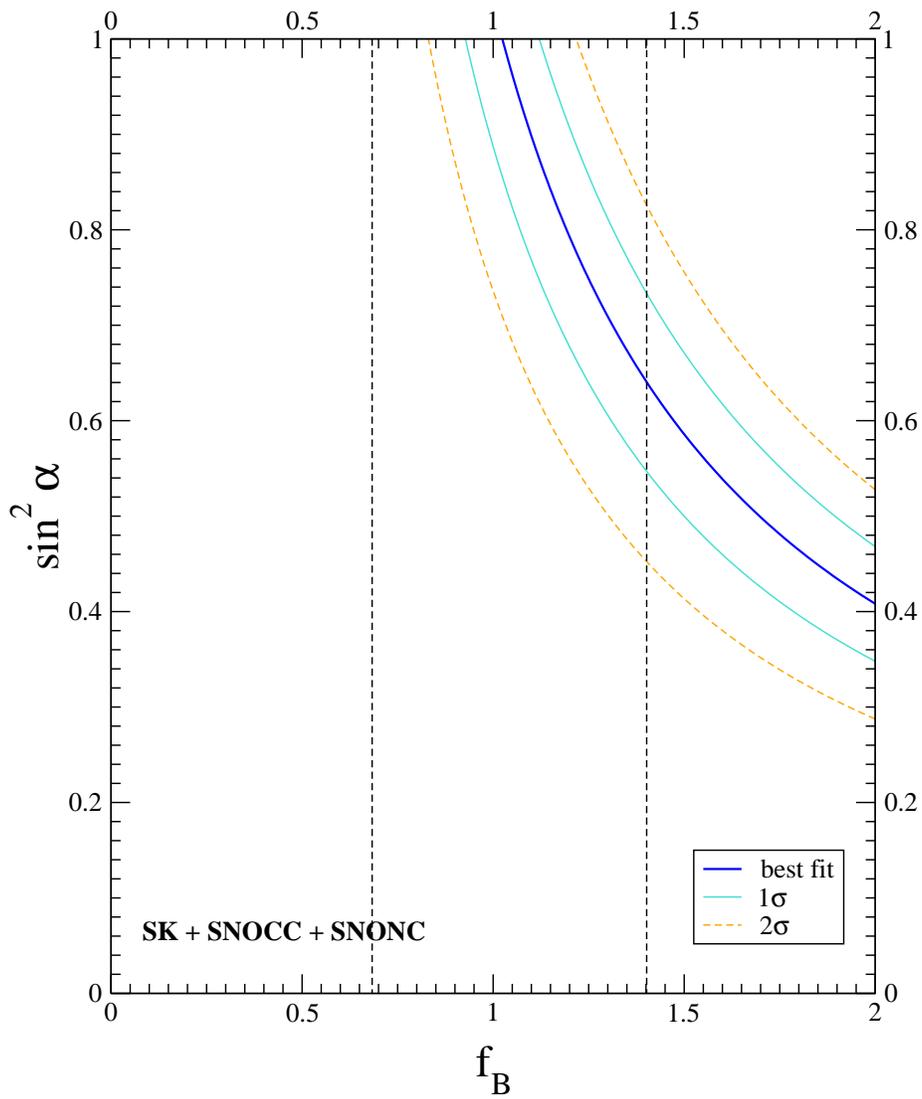}
\label{fig:ncfig2}
\caption
{Best fit value of the $^8B$ neutrino flux $f_B$ shown along with
its $1\sigma$ and $2\sigma$ limits against the model parameter
$\sin^2\alpha$,
representing $\nu_e$ transition into a mixed state ($\nu_a \sin\alpha
+ \nu_s \cos\alpha$). The dashed line denote the $\pm 2\sigma$
limits of the $SSM$. }
\end{center}
\end{figure}

\begin{figure}
\begin{center}
\hspace{10cm}
\epsfxsize=5in
\epsfbox{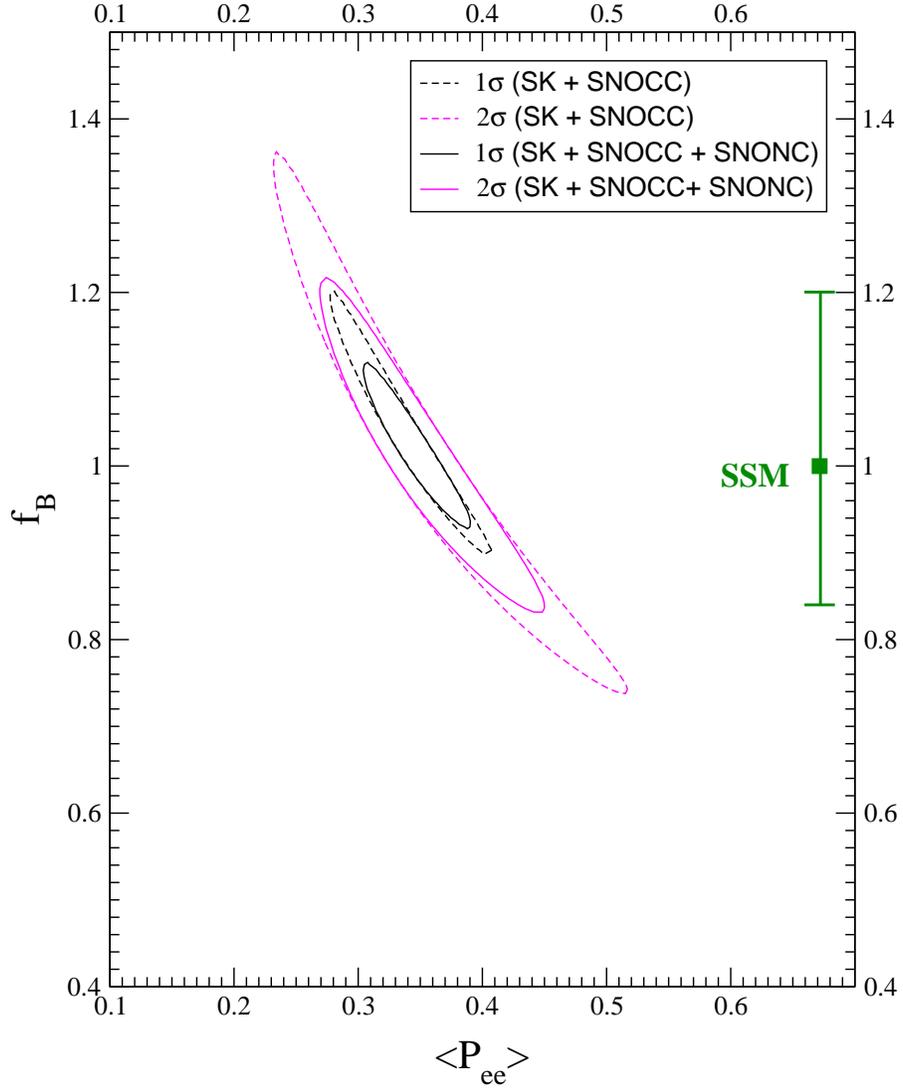}
\label{fig:ncfig3}
\caption{The $1\sigma$ and $2\sigma$ contours of solutions to the $^8B$
neutrino flux $f_B$ and the $\nu_e$ survival probability $P_{ee}$
assuming $\nu_e$ to $\nu_a$ transition.  The 1$\sigma$ $SSM$ error
bar for $f_B$ is indicated on the right.}
\end{center}
\end{figure}

\newpage

\begin{figure}
\begin{center}
\hspace{10cm}
\epsfxsize=5in
\epsfbox{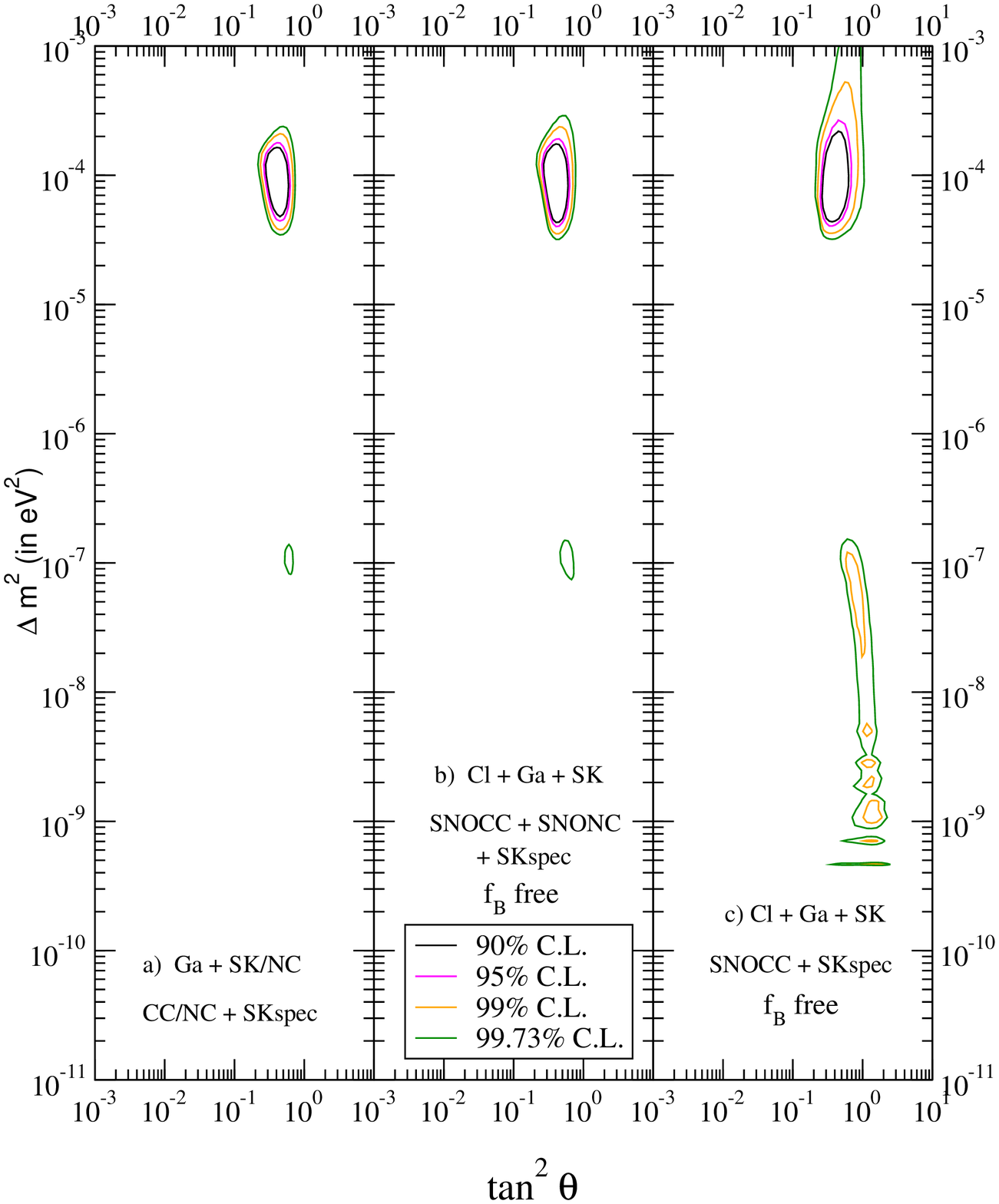}
\label{fig:ncfig5}
\caption{The $\nu_e \rightarrow \nu_a$ oscillation solutions to the
global solar neutrino data using (a) $Ga$ rate, the $SK$ zenith angle 
energy spectra and the $SK$ and $SNO (CC)$ rates, both normalised
to the $SNO (NC)$ rate and (b)
total $Ga$, $Cl$, $SK$, $SNO (CC)$ and $SNO (NC)$ rates along with the
$SK$ zenith angle energy spectra, keeping the $^8B$ flux normalisation
$f_B$ free. 
In both cases we use the $SNO (NC)$ error as the error
in the $^8B$ flux.
The case(c) is similar to (b),
but without using the SNO(NC) rate.}
\end{center}
\end{figure}


\begin{thebibliography}{99}

\bibitem{snonc}
The SNO Collaboration (Q.R. Ahmad {\it et al.}), (to appear in Phys. Rev. 
Lett.), nucl-ex/0204008.

\bibitem{bcgd}A. Bandyopadhyay, S. Choubey, S. Goswami and
D.P. Roy, hep-ph/0203169, (to appear in Mod. Phys. Lett. A).

\bibitem{barger}See {\it e.g.}  V. Barger, D. Marfatia and K. Whisnant,
Phys. Rev. Lett. {\bf 88}, 011302 (2002).

\bibitem{bbp2000}
J.N. Bahcall, M.H. Pinsonneault and S. Basu,
Astrophys. J. {\bf 555}, 990 (2001).

\bibitem{villante}F.L. Villante, G. Fiorentini and E. Lisi,
Phys. Rev. {\bf D59}, 013006 (1999).

\bibitem{lisisno}G.L. Fogli, E. Lisi, D. Montanino and A. Palazzo,
Phys. Rev. {\bf D64}, 093007 (2001).

\bibitem{sk} S. Fukuda {\it et al.,} Super-Kamiokande collaboration,
Phys. Rev. Lett. {\bf 86}, 5651 (2001).  

\bibitem{sno} The SNO Collaboration (Q.R. Ahmad {\it et al.}),
Phys. Rev. Lett. {\bf 87}, 071301 (2001)

\bibitem{snodn}
The SNO Collaboration (Q.R. Ahmad {\it et al.}),
(to appear in Phys. Rev. Lett.),
nucl-ex/0204009.

\bibitem{sg}S. Goswami, D. Majumdar, A. Raychaudhuri, Phys. Rev. {\bf D63},
013003 (2001); ${\it ibid}$ hep-ph/9909453. 
S. Choubey, S. Goswami, N. Gupta and D.P. Roy,
Phys. Rev. {\bf D64}, 053002 (2001);

\bibitem{bcgk}A. Bandyopadhyay, S. Choubey, S. Goswami and
K. Kar, Phys. Lett. {\bf B519}, 83 (2001);
S. Choubey, S. Goswami, K. Kar, H.M. Antia and S.M. Chitre,
Phys. Rev. {\bf D64}, 113001 (2001);
S. Choubey, S. Goswami and D.P. Roy, Phys. Rev. {\bf D65}, 073001 (2002);
A. Bandyopadhyay, S. Choubey, S. Goswami and
K. Kar, Phys. Rev. {\bf D65}, 073031 (2002).

\bibitem{lma}
G.L. Fogli, E. Lisi, D. Montanino, A. Palazzo,
Phys. Rev. {\bf D64}, 093007 (2001);
J.N. Bahcall, M.C. Gonzalez-Garcia, C. Pana-Garay,
JHEP {\bf 0108}, 014 (2001);
P.I. Krastev and A.Yu. Smirnov, e-Print Archive: hep-ph/0108177;
M.V. Garzelli and C. Giunti, JHEP {\bf 0112}, 017 (2001).

\bibitem{ga}
J. N. Abduratshitov et al., SAGE collaboration,
astro-ph/0204245; W. Hampel {\it et al.,} GALLEX
collaboration, Phys. Lett. {\bf B447}, 127 (1999); M. Altman {\it et al.,}
GNO collaboration, Phys. Lett. {\bf B490}, 16 (2000); E. Belloti, talk 
at Gran Sasso National Laboratories, May 17, 2002; T. Kirsten, talk at 
Neutrino 2002, Munich.

\bibitem{cl}
B. Cleveland {\it et al.,}
Ap. J. {\bf 496}, 505 (1998),

\bibitem{smy}M.B. Smy, talk at noon2001, hep-ex/0202020.  

\bibitem{bargernc}V. Barger, D. Marfatia, K. Whisnant and B.P. Wood,
hep-ph/0204253. 

\bibitem{strumia}
P. Creminelli, G. Signorelli and A. Strumia, hep-ph/0102234 
(updated version 3). 

\end{thebibliography}
\end{document}